\documentclass[aps,showpacs, pra,amssymb, amsmath, twocolumn]{revtex4}
\usepackage{amsmath,latexsym,amssymb}
\usepackage{graphicx}
\usepackage{amsmath}
\usepackage{latexsym}
\usepackage{amssymb}
\usepackage{bm}
\usepackage{placeins}
\usepackage{color}

\begin{document}

\title{Coherent control of resonant two-photon transitions by counter-propagating ultrashort pulse pairs}
\author{Woojun Lee, Hyosub Kim, Kyungtae Kim, and Jaewook Ahn}
\email{jwahn@kaist.ac.kr}
\address{Department of Physics, KAIST, Daejeon 305-701, Korea}
\date{\today}

\begin{abstract}
We describe optimized coherent control methods for two-photon transitions in atoms of a ladder-type three-state energy configuration. Our approach is based on the spatial coherent control scheme which utilizes counter-propagating ultrashort laser pulses to produce complex excitation patterns in an extended space. Since coherent control requires constructive interference of constituent transition pathways, applying it to an atomic transition with a specific energy configuration requires specially designed laser pulses. Here, we show, in an experimental demonstration, that the two-photon transition with an intermediate resonant energy state can be coherently controlled and retrieved out from the resonance-induced background, when phase-flipping of the laser spectrum near the resonant intermediate transition is used. A simple reason for this behavior is the fact that the transition amplitude function (to be added to give an overall two-photon transition) changes its sign at the intermediate resonant frequency, thus, by a proper spectral-phase programming, the excitation patterns (or the position-dependent interference of the transition given as a consequence of the spatial coherent control) are well isolated in space along the focal region of the counter-propagating pulses. 
\end{abstract}

\maketitle

\section{Introduction}

Femtosecond laser optics has been widely used, during the last two decades, as a time-resolving spectroscopic means in studying ultrafast time-scale dynamics of variety of quantum systems including atoms, molecules, quasi-particles in solids, etc~\cite{ZewailAC2000, AntoinePRL1996,ChoPRL1990,ChengAPL1991, AhnPRL2001}. Extreme peak intensity of femtosecond laser pulses has also enabled high-order nonlinear optical processes such as multi-photon excitations, high-harmonic generations, and above-threshold ionizations, to list a few~\cite{BrabecRMP2000, FreemanPRL1987, AgostiniRPP2004, CorkumPRL1993}. Besides these usages, coherent control~\cite{Bergmann1998,Shapiro2003,Tanner1985}, the technique of using laser pulse shapes to steer quantum processes towards certain desirable outcomes, has gradually become another important use of femtosecond laser and the information of such laser pulse shapes as obtained through coherent control often plays a crucial role in understanding the quantum structure of the materials under consideration~\cite{BartelsNature2000,HerekNature2002,CruzPNAS2004,ProkhorenkoScience2006}. In that regards, people can furthermore analytically design the laser pulse shapes optimized, although limited to a few quantum systems, for more selective and efficient nonlinear optical processes~\cite{DudovichPRL2001, ClowPRL2008, ReetzPRL2008, LeeOE2011, LimPRA2011a, LimPRA2011b, LeePRA2013}. 

The recent demonstration~\cite{BarmesNP2013} that a counter-propagating pair of ultrafast laser pulses coherently induces Doppler-free two-photon transitions of atoms shows an intriguing possibility towards ultra-precision spectroscopy~\cite{BarmesPRL2013}, especially when being used in conjunction with the femtosecond frequency comb~\cite{HolzwarthPRL2000,JonesScience2000}. This method, termed as spatial coherent control, coherently arranges in time the spectral components of the laser pulse in such a way that all the counter-propagating photon pairs, energy-resonant to the atomic transition ({\it i.e.}, $\hbar\omega_1+\hbar\omega_2=E_e-E_g$), collide only at specific locations along the beam direction. Since the resonance condition varies from one atom species to the another, the atom-specific spectroscopic information can be retrieved by imaging the distinct, if a proper coherent control scheme is used, spatial excitation profile. So far, this powerful method of spatial coherent control has been applied being limited to non-resonant two-photon transitions (or to the cases in which intermediate states are not directly involved with the transition) and yet to resonant two-photon transitions (two-photon transitions with a resonant intermediate state or states). However, there already exist sophisticated pulse-shaping methods~\cite{DudovichPRL2001, LeePRA2013} to deal with the latter case, the resonant two-photon transitions, in the context of optimized single-pulse controls toward enhanced two-photon transitions.

In this paper, we describe an experimental demonstration of the spatial coherent control of the resonant two-photon transition 5S$_{1/2} \rightarrow$ 5P$_{2/3} \rightarrow$ 5D of atomic rubidium ($^{85}$Rb). Ultrashort laser pulses are programmed in such a way that not only (1) counter-propagating photon pairs, and no photon pairs in the same propagation direction, induce the given transition, but also (2) the contributions from the all possible combinations of photon energies, involving such transition, are coherently added for an optimal net transition probability. For this, we first adopt spectral pulse shaping methods: $V$-shape spectral phase programming for the condition (1); and either a spectral phase-step or a spectral phase-window for (2) in the first and second experiments, respectively. 
In the remaining sections, we first theoretically sketch the laser pulse shaping ideas relevant for the resonant and non-resonant two-photon transitions, respectively, in Sec.~II, before the experimental procedure is described in Sec.~III. We then present the experimental result of the spatial coherent control of the two-photon transition of $^{85}$Rb in Sec.~IV and conclude in Sec.~V.

\section{Theoretical Consideration}

We consider a pair of laser pulses, denoted by $\mathcal{E}_1(z,t)$ and $\mathcal{E}_2(z,t)$, respectively propagating along the $\pm z$ directions interacts with a three-level atom in a ladder-type energy configuration.
The two-photon transition amplitude from the ground state $|g\rangle$ to the final state $|e\rangle$ through the intermediate state $|i\rangle$ is then obtained from the second-order Dyson series in the perturbative interaction regime as
\begin{eqnarray}
c_{eg}(z,t)&=&-\frac{\mu_{ei}\mu_{ig}}{2 \pi \hbar^2} \int_{-\infty}^t d t_1   e^{i \omega_{ei} t_1} \mathcal{E}(z,t_1) \nonumber  \\
&& \times \int_{-\infty}^{t_1} d t_2   e^{i \omega_{ig} t_2} \mathcal{E}(z,t_2),
\end{eqnarray}
where $\mathcal{E}(z,t)= \mathcal{E}_1(z,t) + \mathcal{E}_2(z,t)$ and $\mu_{ei}$ and $\mu_{ig}$ are the corresponding dipole moments. The integration provides the transition amplitude in terms of the spectral amplitudes (including phase) $E_1(\omega)=FT[\mathcal{E}_1(0,t)]$ and $E_2(\omega)=FT[\mathcal{E}_2(0,t)]$ by
\begin{eqnarray}
c_{eg} (z) &=& i \frac{\mu_{ei}\mu_{ig}}{\hbar^2} \bigg[ i \pi E(\omega_{ig}) E(\omega_{ei}) 
\nonumber \\ && + \int_{-\infty}^\infty \frac{E(\omega) E(\omega_{eg}-\omega)}{\omega_{ig}-\omega} \bigg] ,
\label{ceg}
\end{eqnarray}
where $E(\omega) = E_1 (\omega) e^{-i \omega {z}/{c}} + E_2 (\omega) e^{i \omega {z}/{c}}$ and $t\rightarrow \infty$ is assumed. We denote the first term in the bracket of Eq.~\eqref{ceg} by $c_{r} (z)$, the resonant two-photon transition, and the second term by $c_{nr} (z)$, the non-resonant two-photon transition, {\it i.e.}, 
\begin{equation}
c_{eg} (z) = c_{r} (z) + c_{nr} (z).
\end{equation}   
which can be respectively written down in terms of $E_1(\omega)$ and $E_2(\omega)$ as  
\begin{widetext}
\begin{eqnarray}
c_{r}(z) &=& - \pi \frac{\mu_{ei}\mu_{ig}}{\hbar^2} \big[ E_1 (\omega_{ig}) E_1 (\omega_{ei}) + E_2(\omega_{ig}) E_2(\omega_{ei}) e^{2i \omega_{eg} {z}/{c}} \nonumber \label{cncnr1} \\
&& \quad\quad\quad\quad\quad +  E_1 (\omega_{ig}) E_2 (\omega_{ei}) e^{2i \omega_{ei} {z}/{c}} + E_2 (\omega_{ig}) E_1(\omega_{ei}) e^{2i \omega_{ig} {z}/{c}} \big], \label{cn} \\
c_{nr}(z) &=& i \frac{\mu_{ei}\mu_{ig}}{\hbar^2} \int_{-\infty}^\infty \frac{d\omega}{\omega_{ig}-\omega} \big[ E_1 (\omega) E_1 (\omega_{eg}-\omega) + E_2(\omega) E_2(\omega_{eg} -\omega) e^{2i \omega_{eg} {z}/{c}}  \nonumber \\
&& \quad\quad\quad\quad\quad\quad\quad\quad\quad + E_1 (\omega) E_2 (\omega_{eg} - \omega) e^{2 i (\omega_{eg}-\omega) {z}/{c}} + E_2 (\omega) E_1(\omega_{eg}-\omega) e^{2i \omega {z}/{c}} \big].
\label{cncnr2}
\end{eqnarray}
\end{widetext}

Note that the first two terms in each bracket in Eqs.~\eqref{cncnr1} and \eqref{cncnr2} are the single-pulse contributions from either direction of the pulse propagations, and the last two terms are caused by the presence of both pulses. Furthermore, the all terms in $c_{r} (z)$ and the first two terms (the single-pulse contribution) in $c_{nr} (z)$ have the position-dependent phase factors (1, $e^{2i\omega_{eg}z/c}$, $e^{2i\omega_{ei}z/c}$, $e^{2i\omega_{ig}z/c}$, etc.) that are not directly coupled with the laser spectrum $\omega$, while the last two terms (the non-resonant transition contribution by the presence of the both pulses) in $c_{nr} (z)$ has explicit correlation between the position coordinate $z$ and the laser spectrum $\omega$, which means that the last two terms in  $c_{nr} (z)$, and no other terms, contribute to the excitation profile. 

The position dependent part of the excitation spatial profile is approximately given by
\begin{widetext}
\begin{equation}
S(z) \propto \left| \int^\infty_{-\infty}   \frac{d\omega {(\omega_{ei}-\omega_{ig})}
A(\omega)A(\omega_{eg}-\omega)}{(\omega_{ig}-\omega)(\omega_{ei}-\omega)}
e^{i[\Phi(\omega)+\Phi(\omega_{eg}-\omega)+2i\omega z/c]} \right|^2,
\label{Sz}
\end{equation}
\end{widetext}
where the spectral amplitude function $A(\omega)$ and the spectral phase function $\Phi(\omega)$ are defined the same for the both pulses, when they are split from a single laser pulse, as
\begin{equation}
E_1(\omega)=E_2(\omega)=A(\omega)e^{i\Phi(\omega)}.
\end{equation}
Therefore, in the context of the spatial coherent control, the phase function $\Phi(\omega)$ can be programmed in such a way that the two-photon transition components in the integral calculation in Eq.~\eqref{Sz} satisfy a constructive interference condition only at specific positions, so that the excitation spatial profile $S(z)$ can be used as a spectroscopic means.

\section{Experimental Description}
\begin{figure}[tbh]
    \centerline{\includegraphics[width=0.45\textwidth]{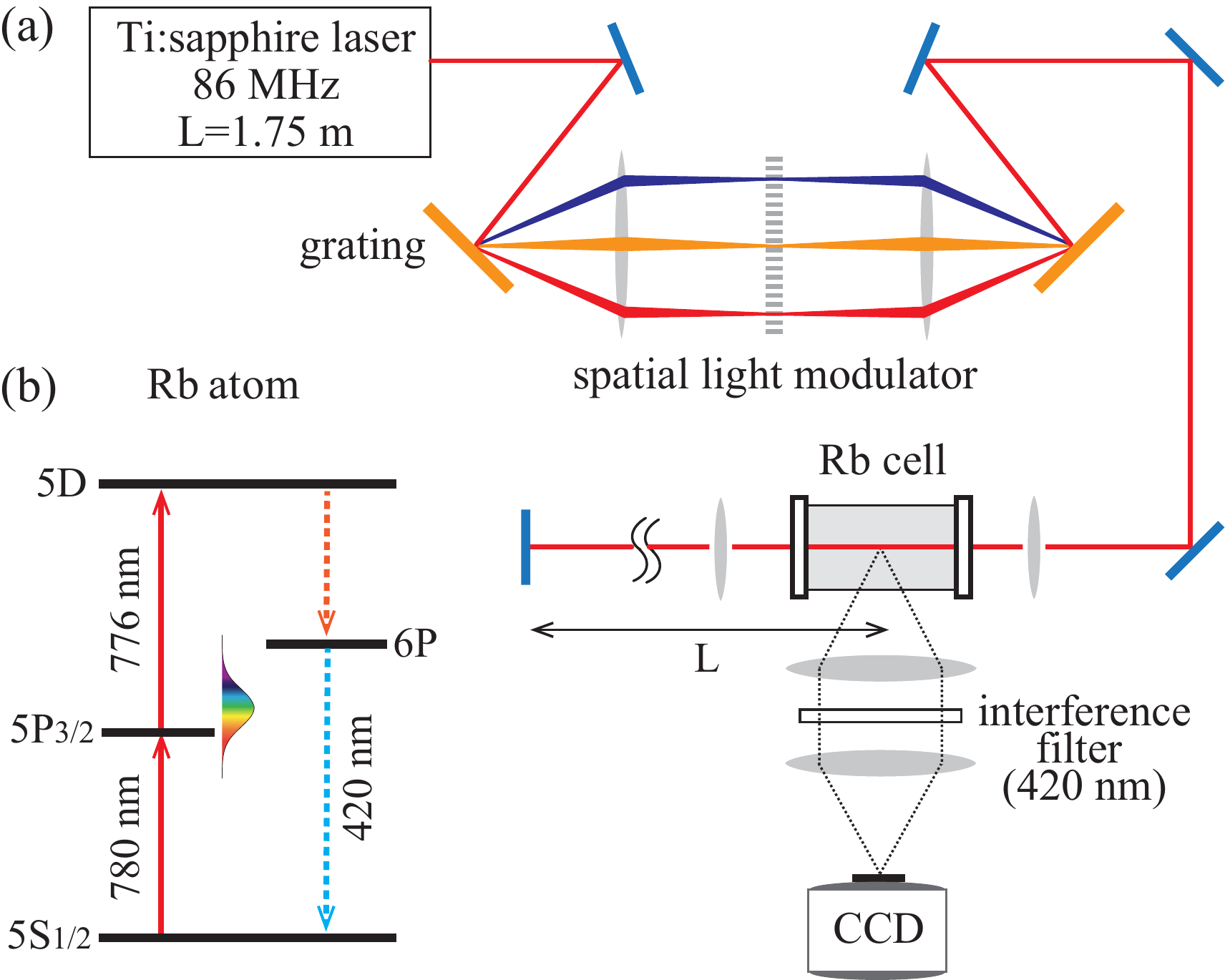}}
    \caption{(Color online) (a) Schematic diagram of the experimental setup. Ultrafast laser pulses were programmed by a spatial light modulator to interact with rubidium atoms in a colliding pulse geometry. The cavity length $L=1.75$~m of the laser was matched with the extra travel of the reflected pulses. (b) The energy level diagram of atomic rubidium. Atoms are excited from 5S$_{1/2}$ to 5D and decayed to 6P. The fluorescence from 6P was monitored.} 
    \label{FigS3}
\end{figure}

The experimental setup is schematically shown in Fig.~\ref{FigS3}(a). Femtosecond optical pulses were generated from a conventional Ti:sapphire laser oscillator (of a cavity length $L=1.75$~m) mode-locked around the center frequency set to the two-photon transition frequency ({\it i.e.} $\omega_{o}=\omega_{eg}/2$). The laser bandwidth (FWHM) was 25~nm in wavelength scale. The pulses were then spectrally resolved in a $4f$-geometry Fourier plane and phase-modulated by a 128-pixel spatial-light modulator (SLM)~\cite{WeinerRSI2000} to program $\Phi(\omega)$. The focal length of the $4f$-geometry was 200~mm and the groove number of both gratings was 1200~mm$^{-1}$~\cite{Martinez}. The width of each pixel in SLM was $100~\mu$m and the pulse bandwidth incident into each pixel was 0.37~nm in wavelength scale. After pulse-shaping, the pulses were focused in a rubidium vapor cell ($^{85}$Rb) located at $z=0$, de-focused, collimated, and then reflected back by a mirror located at $z=L$, so that each counter-propagating laser pulse collided with the next pulse in the vapor cell. The focal length of both focusing lenses was 200~mm. The beam diameter and the Rayleigh range of the focus were 100~$\mu$m and 10~mm, respectively.

The $^{85}$Rb atoms were two-photon excited from the ground state $|g\rangle$=5S$_{1/2}$ to the final state $|e\rangle$=5D via the intermediate state $|i\rangle$=5P$_{3/2}$ [see Fig. 1(b)]. The corresponding frequencies (wavelengths) were $\omega_{ig}/2\pi=384.6$~ THz ($\lambda_{ig}=780$~nm), $\omega_{ei}/2\pi=386.6$~THz ($\lambda_{ei}=776$~nm), and $\omega_o/2\pi=385.6$~THz ($\lambda_{o}=778$~nm), respectively~\cite{RbNist}. Note that the D1 transition to 5P$_{1/2}$ state was out of the laser spectral range. The excited atoms in the 5D were first decayed to 6P, and then the spatial profile of the fluorescence at 420~nm (6P $\rightarrow$ 5S$_{1/2}$) was imaged by a CCD camera through one-to-one telescope imaging by a pair of $f=25$~mm lenses. The image resolution of the camera was 4.54~$\mu$m.

\section{Results and Discussion}
\begin{figure}[tbh]
    \centerline{\includegraphics[width=0.4\textwidth]{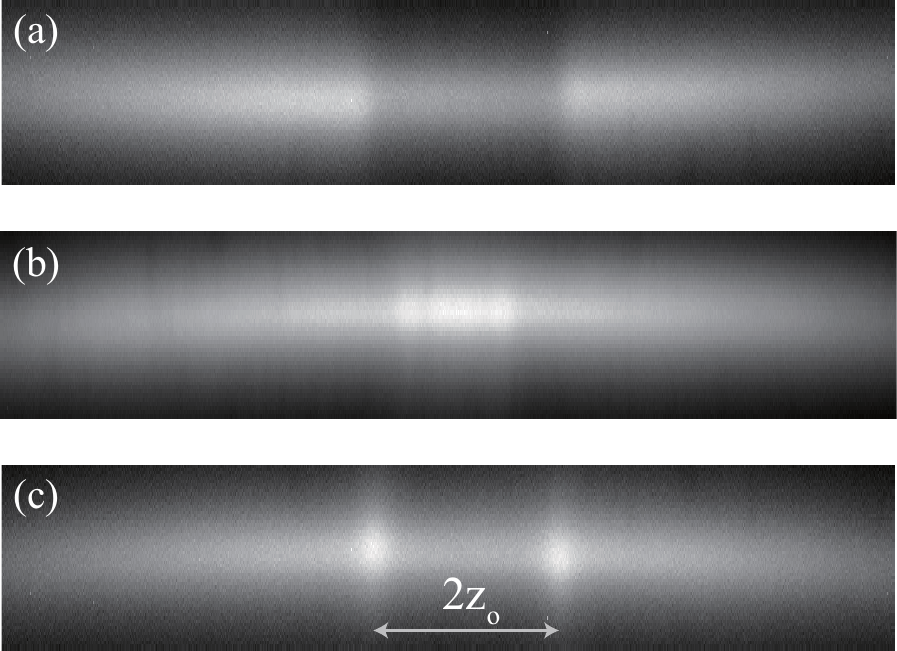}}
    \caption{(Color online) Photo images of the fluorescence signals: (a,b) $V$-shape spectral phase functions $\Phi_V(\omega)$ are used respectively with (a) $\alpha=-1.4$~ps, (b) $\alpha=+1.0$~ps. (c) The spectral phase function with a phase step $\Phi_1(\omega)$ with $\beta= \pi$ and $\alpha=-1.4$~ps. [$\Phi_V(\omega)$ and $\Phi_1(\omega)$ are defined in Eqs.~\eqref{eq_phiV} and \eqref{eq_phi1}, respectively, and $z_o=420$~$\mu$m in (c).]} 
    \label{Photo_both}
\end{figure}

Before the result of the resonant-two-photon experiments is presented, we first consider the {\it non-resonant two-photon transition} case, which is the case when the laser spectrum does not cover the transition to the intermediate state $|i\rangle$ ({\it i.e.}, $\omega_{ig} < \omega_{\rm min}$ for $\omega_{ig}<\omega_{ei}$). The $V$-shape spectral phase function, defined in Ref.~\cite{BarmesNP2013} as
\begin{equation}
\Phi_V(\omega)=\alpha|\omega-\omega_o|,
\label{eq_phiV}
\end{equation}
makes two distinct local functions in the spatial excitation profile $S(z)$ in Eq.~\eqref{Sz}, which are conceptually given by
\begin{equation}
S(z) \propto \delta(z-z_o) + \delta(z+z_o), 
\label{eq_afg2}
\end{equation}
where $z_o=\alpha c$  is a constant determined by the phase slope $\alpha=|d\Phi_V/d\omega|$ and $\delta(z)$ is a Dirac delta function. However, when this phase function $\Phi_V(\omega)$ is used for the {\it resonant two-photon} case, the image appears as in Fig.~\ref{Photo_both}(a), when $\alpha<0$ is applied. The darker region in $z<|z_o|$ violates the causality for the resonant two-photon transition, because the $|g\rangle \rightarrow |i\rangle$ occurs later in time than $|i\rangle \rightarrow |e\rangle$. When $\alpha>0$,  on the other hand, a brighter region appears in $z<|z_o|$ as shown in Fig.~\ref{Photo_both}(b) but the given excitation profile is not spatially localized.  

In order to implement spatial coherent controls useful for {\it resonant two-photon transitions}, alternative phase functions should be considered. In this case, the intermediate resonant state $|i\rangle$ is located within the laser spectrum  ({\it i.e.}, $\omega_{\rm min} < \omega_{ig} < \omega_{\rm max}$ directly involved with the two-photon transition.  For example, the image shown in Fig.~\ref{Photo_both}(c) was taken with the following phase function:
\begin{equation}
\Phi_1(\omega)=\alpha|\omega-\omega_o|+\beta\Theta(\omega-\omega_{ig}),
\label{eq_phi1}
\end{equation}
with the Heaviside step function defined by $\Theta(x)=1$ for $x\ge 0$ and 0 for $x<0$. The physics behind this behavior is from  the dispersion term $1/(\omega_{ig}-\omega)$ in Eq.~\eqref{Sz}, which causes a phase flip across the resonant intermediate-state frequency and results in a destructive interference among the two-photon transition pathways in the given spectral sum. However, the second phase-step term $\beta\Theta(\omega-\omega_{ig})$ in Eq.~\eqref{eq_phi1} recovers the constructive interference. The $\pi$-step phase function, or a similar concept, has been tested before in simple coherent control experiments of two-photon transitions with one or two resonant intermediate states~\cite{DudovichPRL2001, LeePRA2013}.

\begin{figure}[tbh]
    \centerline{\includegraphics[width=0.5\textwidth]{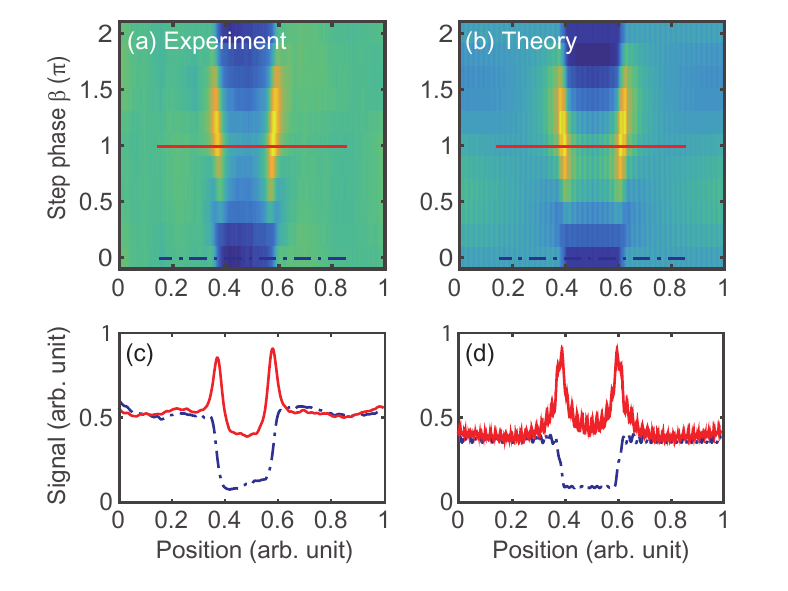}}
    \caption{(Color online) The first phase shaping experiment: (a, c) spatial fluorescence signals for $\Phi_1(\omega)$ ($V$-shape + phase step) spectral functions, (b,d) numerical simulation. Signals at $\beta=0$ (blue dash-dot line) and $\pi$ (red line) are extracted from (a) and (b) to show in (c) and (d), respectively. [Unwanted background signals due to the position-insensitive resonant two-photon transitions are subtracted for clarity by means of numerical fitting.]} 
    \label{Fig3}
\end{figure}

Figure~\ref{Fig3} shows the result of the first experiment performed with the spectral phase function $\Phi_1(\omega)$, in comparison with a corresponding numerical calculation. By maintaining the phase slope value fixed at $\alpha=-1.4$~ps, the phase step value was changed from $\beta=0$ to $2\pi$, and the fluorescence signals (measured along the $z$ direction where the position is shown in the horizontal axis) were plotted as a function of $\beta$ (along the vertical axis) in Fig.~\ref{Fig3}(a). 
As expected, sharp fluorescence peaks appear at $z=z_o$ and $-z_o$, when $\beta=\pi$ is applied. The spatial  excitation profiles measured at $\beta=0$ (blue dash-dot line) and $\pi$ (red line) are respectively shown in Fig.~\ref{Fig3}(c). This experimental result agrees well with the numerical simulation as shown in Figs.~\ref{Fig3}(b) and \ref{Fig3}(d). 

\begin{figure}[tbh]
    \centerline{\includegraphics[width=0.5\textwidth]{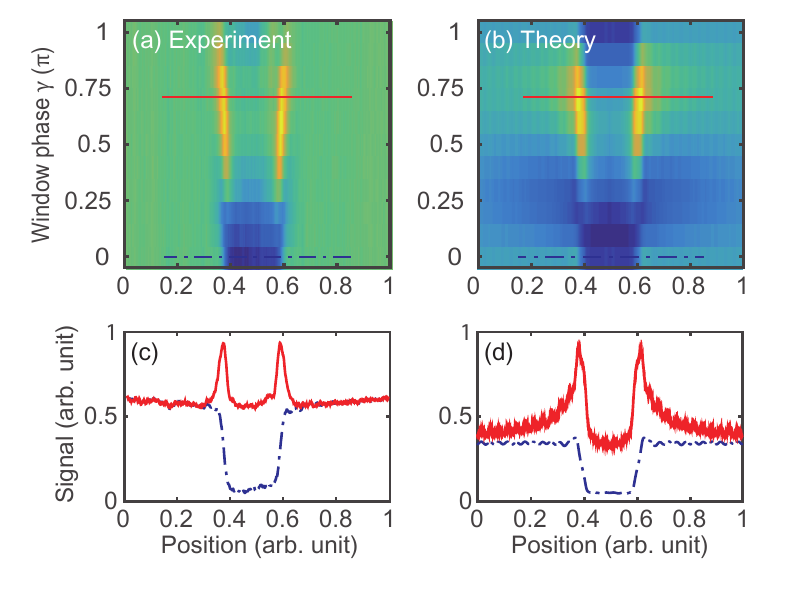}}
    \caption{(Color online) The second phase shaping experiment: (a, c) spatial fluorescence signals for $\Phi_2(\omega)$ ($V$-shape + phase block) spectral functions, (b,d) numerical simulation. Signals at $\gamma=0$ (blue dash-dot line) and $0.65\pi$ (red line) are extracted from (a) and (b) to show in (c) and (d), respectively.} 
    \label{Fig4}
\end{figure}

In the second experiment shown in Fig.~\ref{Fig4}, the spectral block between $\omega_{ig}$ and $\omega_{eg}-\omega_{ig}$ was phase-rotated from $\gamma=0$ to $\pi$, {\it i.e.}, 
the spectral phase function is given by
\begin{equation}
\Phi_2(\omega)=\alpha|\omega-\omega_o|+\gamma[\Theta(\omega-\omega_{ig})-\Theta(\omega-\omega_{eg}+\omega_{ig})],
\end{equation}
where  the same $\alpha=-1.4$~ps was used as in the first experiment. This type of phase modulation is theoretically equivalent to the one in the first experiment, when $\gamma=\beta/2$~\cite{LeePRA2013}. Both the $\Phi_1(\omega)$ in the first experiment and $\Phi_2(\omega)$ with $\gamma=\pi+\beta/2$ flip the sign of the dispersion term $1/(\omega_{ig}-\omega)(\omega_{ei}-\omega)$ in Eq.~\eqref{Sz} in the spectral region $\omega_{ig}<\omega<\omega_{ei}$ with respect to the rest of the spectral region so that transitions from the whole spectral region become in phase. The result shown in Figs.~\ref{Fig4}(a) and \ref{Fig4}(c) clearly shows the unresolved spatial profile at $\gamma=0$ gradually becomes resolved as $\gamma \rightarrow \pi/2$. The maximal visibility is achieved at $\gamma=\pi/2$ in a good agreement with the theoretical prediction in Figs.~\ref{Fig4}(b) and \ref{Fig4}(d). 

The maximum peak signal in Fig.~\ref{Fig3} (Fig.~\ref{Fig4}) appears at $\beta=1.1\pi$ ($\gamma=0.65\pi$), rather than $\beta=\pi$ ($\gamma=\pi/2$). This behavior is accounted for by the presence of the resonant amplitude term $c_r(z)$ in Eq.~\eqref{cncnr1} as follows. If we divide the non-resonant term $c_{nr}(z)$ at $z=\pm z_0$ into $c_{nr}^+$ and $c_{nr}^-$, where $c_{nr}^+$ and $c_{nr}^-$  denote the positive and negative parts of $c_r(z)$ integrated for $\omega>\omega_{ig}$ and $\omega<\omega_{ig}$, respectively, the $c_{nr}^+$ term is 90$^\circ$ phase-advanced and the $c_{nr}^-$ term is 90$^\circ$ phase-delayed with respect to the $c_r$. Since the phase of the $c_r$ term was rotated with the $c_{nr}^+$ term, due to the limited spectral resolution of the pulse-shaper, both of them phase-rotated together in the complex plane by $\beta$ ($2\gamma$). Therefore, the maximum transition, or the phase-overlap of the all three terms, occurs at $\beta=\pi+\theta_o$ ($2\gamma=\pi+\theta_o$), where  $\theta_o=\tan^{-1}(c_r/c_{nr}^+)$. The result in Fig.~\ref{Fig3} (Fig.~\ref{Fig4}) measures  $\theta_o=0.1\pi$ ($0.3\pi$) and our calculation estimates the ratio between the amplitudes $c_r : c_{nr}^+ = 1:3$. Therefore, the estimated result is consistent with the measured phase shift $\Delta \beta=0.1\pi$ ($\Delta \gamma=0.3\pi$).

\section{Conclusion}
In summary, we have performed coherent control experiments with counter-propagating phase-modulated ultrashort pulses for
the two-photon excitations in the ladder-type three-state quantum system of atomic rubidium. The laser pulses designed based on the spatial coherent control scheme to phase-flip the laser spectrum near the resonant intermediate transition have successfully produced spatially localized excitation patterns out from the resonance-induced background along the focal region of the counter-propagating pulses. The resulting constructive interference phenomena among the constitute non-resonant two-photon transitions are theoretically confirmed and verified by numerical calculations.

\begin{acknowledgements}
\end{acknowledgements}


\begin{thebibliography}{99}

\bibitem{ZewailAC2000} A. H. Zewail, 
Angew. Chem. Int. Ed. Engl. {\bf 39}, 2587 (2000).
\bibitem{AntoinePRL1996} P. Antoine, A. L'Huillier, and M. Lewenstein, 
Phys. Rev. Lett. {\bf 77},1234 (1996).
\bibitem{AhnPRL2001} J. Ahn, D. N. Hutchinson, C. Rangan, and P. H. Bucksbaum, 
Phys. Rev. Lett. {\bf 86}, 1179 (2001).

\bibitem{ChoPRL1990} G. C. Cho, W. Kutt, and H. Kurz, 
Phys. Rev. Lett. {\bf 65}, 764 (1990).
\bibitem{ChengAPL1991} T. K. Cheng, S. Vidal, M. J. Zeiger, G. Dresselhaus, M. S. Dresselhaus, and E. P. Ippen, 
Appl. Phys. Lett. {\bf 59}, 1923 (1991).

\bibitem{BrabecRMP2000} T. Brabec and F. Krausz, 
Rev. Mod. Phys. {\bf 72}, 545 (2000).
\bibitem{FreemanPRL1987} R. R. Freeman, P. H. Bucksbaum, H. Milchberg, S. Darack, D. Schumacher, and M. E. Geusic,
Phys. Rev. Lett. {\bf 59}, 1092 (1987).
\bibitem{CorkumPRL1993} P. B. Corkum,
Phys. Rev. Lett. {\bf 71}, 1994 (1993).
\bibitem{AgostiniRPP2004} P. Agostini and L. F. DiMauro,
Rep. Prog. Phys. {\bf 67}, 813 (2004).

\bibitem{Bergmann1998} K. Bergmann, H. Theuer, and B. W. Shore,
Rev. Mod. Phys.~{\bf 70}, 1003 (1998).
\bibitem{Shapiro2003} M. Shapiro and P. Brumer, {\it Principles of the quantum control of molecular processes},
(Wiley, New York, 2003).

\bibitem{Tanner1985} D.J. Tanner and S. A. Rice, 
J. Chem. Phys.~{\bf 83}, 5013 (1985).

\bibitem{BartelsNature2000} R. Bartels, S. Backus, E. Zeek, L. Misoguti, G. Vdovin, I. P. Christov, M. M. Murnane, and H. C. Kapteyn, 
Nature {\bf 406}, 164 (2000).
\bibitem{HerekNature2002} J. L. Herek, W. Wohlleben, R. J. Cogdell, D, Zeidler, and M. Motzkus, 
Nature {\bf 417}, 533–535 (2002).
\bibitem{CruzPNAS2004} J. M. Dela Cruz, I.  Pastirk, M. Comstock, V. V. Lozovoy, and M.  Dantus,
Proc. Natl. Acad. Sci. {\bf 101}, 16996 (2004).
\bibitem{ProkhorenkoScience2006}
V. I. Prokhorenko, A. M. Nagy, S. A. Waschuk, L. S. Brown, R. R. Birge, and R. J. Dwayne Miller,
Science {\bf 313}, 1257 (2006).

\bibitem{DudovichPRL2001} N. Dudovich, B. Dayan, S. M. GallagherFaeder, and Y. Silberberg, 
Phys. Rev. Lett. {\bf 86}, 47 (2001).
\bibitem{ClowPRL2008} S. D. Clow, C. Trallero-Herrero, T. Bergeman, and T. Weinacht,
Phys. Rev. Lett. \textbf{100}, 233603 (2008).
\bibitem{ReetzPRL2008} M. Reetz-Lamour, T. Amthor, J. Deiglmayr, and M. Weidem\"{u}ller, 
Phys. Rev. Lett. {\bf 100}, 253001 (2008).
\bibitem{LeeOE2011} S. Lee, J. Lim, C. Y. Park, and J. Ahn,
Opt. Express {\bf 19}, 2266 (2011).   
\bibitem{LimPRA2011a} J. Lim, H. G. Lee, J. U. Kim, S. Lee, and J. Ahn,
Phys. Rev. A \textbf{83}, 053429 (2011).
\bibitem{LimPRA2011b} J. Lim, H. G. Lee, S. Lee, and J. Ahn,
``Quantum control in two-dimensional Fourier transform spectroscopy,'' 
Phys. Rev. A \textbf{84}, 013425 (2011).
\bibitem{LeePRA2013} H. G. Lee, H. Kim, J. Lim, and J. Ahn, 
Phys. Rev. A {\bf 88}, 053427 (2013).

\bibitem{BarmesNP2013} I. Barmes, S. Witte, and K. S. E. Eikema, 
Nat. Photonics {\bf 7}, 38 (2013).
\bibitem{BarmesPRL2013} I. Barmes, S. Witte, and K. S. E. Eikema, 
Phys. Rev. Lett. {\bf 111}, 023007 (2013).

\bibitem{HolzwarthPRL2000}
R. Holzwarth, Th. Udem, T. W. H\"{a}nsch, J. C. Knight, W. J. Wadsworth, and P. St. J. Russell,
Phys. Rev. Lett. {\bf 85}, 2264 (2000).
\bibitem{JonesScience2000}
D. J. Jones, S. A. Diddams, J. K. Ranka, A. Stentz, R. S. Windeler, J. L. Hall, and S. T. Cundiff,
Science {\bf 288}, 635 (2000).

\bibitem{WeinerRSI2000} A. M. Weiner, 
Rev. Sci. Inst. {\bf 71}, 1929 (2000).

\bibitem{Martinez} R. L. Fork, O. E. Martinez, and J. P. Gordon, 
Opt. Lett. {\bf 9}, 150 (1984).

\bibitem{RbNist} J. E. Sansonetti, %
J. Phys. Chem. Ref. Data, {\bf 35}, 301 (2006).

\end{thebibliography}
\end{document}